\documentclass{iaus}
\usepackage{graphicx}

\title[Dark galaxies as XMD galaxy progenitors] 
{``Dark Galaxies'' and Local Very Metal-Poor Gas-Rich Glaxies: Possible
Interrelations
}
\author[S.A. Pustilnik]   
{Simon A. Pustilnik}%
\affiliation{Special Astrophysical Observatory, Russian Academy of
Sciences, Nizhnij Arkhyz, Karachai-Circassia, 369167, Russia
\break email: sap@sao.ru\\[\affilskip]
}
\pubyear{2008}
\volume{244}  
\pagerange{417--421}
\date{?? and in revised form ??}
\jname{Dark Galaxies and Lost Baryons}
\editors{J. I. Davies \& M. J. Disney, eds.}

\begin{document}

\maketitle

\begin{abstract}
There are only a few ``dark galaxy'' candidates discovered to date in the
local Universe. One of the most prominent of them is the SW component of a
merging system HI 1225+01. On the other hand, the number of known very
metal-poor gas-rich dwarfs similar to I~Zw~18 and SBS~0335--052~E,W has
grown drastically during the last decade, from a dozen and a half to about
five dozen. Many of them are very gas-rich, having from $\sim$90 to 99~\% of
all baryons in gas. For some of such objects that have the deep
photometry data, no evidences for the light of old stars are found. At least a
half of such galaxies with the prominent starbursts have various evidences
of interactions, including advanced mergers. This suggests that a fraction of
this group objects can be a kind of very stable protogalaxies (or ``dark
galaxies''), which have recently experienced strong disturbances from nearby
massive galaxy-size bodies. Such a collision caused the gas instabilities and
its collapse with the subsequent onset of starburst. We briefly discuss the
morphology and gas kinematics for the subsample of the most metal-poor dwarfs
that illustrate this picture. We discuss also the relation of these rare
galaxies to the processes by which ``dark galaxies'' can occasionally
transform to optically visible galaxies.
\end{abstract}

\firstsection 

\section{Introduction}

The possible existence of a large number of dark galaxies was suggested to
reconcile the drastic difference between the predicted number density of
low-mass Cold DM halos and the observed number density of low-mass galaxies
(e.g., \cite{Klypin99, Moore99}). Despite the fact that the gap diminished
significantly
during the last years thanks to discovery of many very low-mass galaxies
in the Local Group (e.g., Moore, this meeting), it is still too large. The
observational methods of searching for dark galaxies were discussed, e.g., by
\cite{Trentham01}.
The range of global parameters, in which one expects the dark galaxies to
exist, was explored in \cite{Verde02}, \cite{TW05} and \cite{Davies06}
based on one of CDM models of disk galaxy formation by \cite{Mo98}.
While the predictions of the original version of \cite{TW05} work
differed strongly from those of \cite{Davies06}, the revised analysis by
E.~Taylor (this meeting) resulted in similar ranges of dark galaxy global
parameters.

\section{Dark Galaxy candidates vs model-predicted objects}

There are a few Dark Galaxy candidates known to date. We summarize some of
their observational parameters in Table~\ref{tab:dark}. Three of these four
objects were discussed during this Symposium in the talks given by M.~Haynes,
J.~Davies and E.~Brinks. No optical counterparts were found for any of them.
The upper limits on their central surface brightness (SB) are at the levels
of 27--27.5 $V$-mag sq.arcsec$^{-2}$ (\cite{Salzer91}, \cite{Minchin05},
\cite{Walter05}), except HI J0325--3655 near FCC 35, for which no optical
counterpart is visible on the SERC images (\cite{Putman98}).
The distance-dependent parameters for HI 1225+01~SW are given for two
possible distance extremes of 10 and 20 Mpc. V$_{\rm rot}$ is estimated
either from the velocity fields, or from the maximal velocity widths. The
first glimpse on the table reveals a rather large scatter of main parameters
for these candidates. In addition, the new data presented
at the Symposium, suggests that HIJASS J1021+6824 along with many smaller
HI features can have the tidal origin due to the strong interactions in M~81
group (Brinks, this meeting).
Moreover, according to the ALFALFA HI map, VIRGOHI~21 looks like a part of a
long tail stretching from the massive spiral NGC~4254 (R.~Giovanelli, this
meeting) which is probably an evidence against its dark galaxy nature.

On the other hand, there are $\Lambda$CDM N-body simulations mentioned above,
predicting the regions of the parametric space of baryon aggregates in which
one can expect to find dark galaxies, defined as baryon ``disks'' inside DM
halos with no stars formed so far. Therefore, it is reasonable to compare
these predicted properties with those of ``dark galaxy'' candidates and of
some their possible descendants.
Since as told in the Introduction, both \cite{Davies06} and Taylor (2007, this
meeting) give in general the consistent ranges of main parameters of dark
galaxies, we base our further on the published results of \cite{Davies06}.
They can be summarized as follows. The total range of M(HI) is of 10$^5$ to
10$^9$ M$_{\odot}$, with
only $\sim$2\% of simulated objects to have M(HI)=10$^8$-10$^9$~M$_{\odot}$.
HI column densities vary in the range of (0.5--5)$\times$10$^{20}$
at.~cm$^{-2}$. The range of V$_{\rm rot}$ is of 5 to 70 km~s$^{-1}$,
with less than $\sim$1\% to have V$_{\rm rot}$ $>$ 30 km~s$^{-1}$.
The accepted baryon mass fraction (M$_{\rm bar}$/M$_{\rm tot}$)
according to theoretical expectations varies between 0.01 and 0.05.

\begin{table}\def~{\hphantom{0}}
  \begin{center}
  \caption{Main parameters of candidate dark galaxies}
  \label{tab:dark}
  \begin{tabular}{lccccc}\hline
Name              & M(HI)$^1$& V$_{\rm rot}$$^2$ & M$_{\rm bar}$/M$_{\rm tot}$ &HI col.dens.$^3$ & Lin.size$^4$ \\\hline
HI 1225+01~SW     & 2.8-11.2 & ~~13  & {\bf 0.4-1.0} & $<$3-4  & 17-34   \\
VIRGOHI~21        & 0.4      & ~100  & {\bf 0.002}   & 0.3     & 14     \\
HIJASS J1021+6824 & 1.5      & ~~40  & 0.036     & 1.8     & 30     \\
HI J0325--3655    & 2.2      & $<$20 & $>$0.1    & 0.2     & 16     \\ \hline
  \end{tabular}
 \end{center}
$^{1}$ in units 10$^8$~M$_{\odot}$, $^{2}$ in units km~s$^{-1}$,
$^3$ in units 10$^{20}$ at.~cm$^{-2}$, $^{4}$ in kpc.
\end{table}

Comparing the observed properties of candidate ``dark galaxies'' with those
predicted from the N-body simulations, one concludes that they do not match
each other well. The observed HI masses are too large. Could this be happening
partially
due to a selection effect? Or are we (mainly) dealing with ``wrong''
candidates? On the other hand, accounting for possible descendants of dark
galaxies among XMD galaxies (see Sect.~\ref{XMD_merger}), many of which are
also ``massive'', are we that confident about the predicted mass range of
dark galaxies?

If VIRGOHI~21 is a dark galaxy, its
M$_{\rm bar}$/M$_{\rm tot}$=0.002 appears to be atypically small. On the other
hand, if HI 1225+01~SW ($\sim$0.5),  is a real dark galaxy, this parameter
is also a challenge for commonly-accepted models. Can such a high baryon
mass-fraction be observed only in transient entities (e.g., ``tidal dwarfs'')
characteristic of interacting systems?
If one of new XMD BCGs (see Table~\ref{XMD_merger}) and two new HI-rich
isolated
dwarf galaxies with similar baryon mass-fractions found in ALFALFA
(M.~Haynes, this meeting) are real, they probably prove that such rare objects
can exist as stable aggregates. If this is true, then we can say that
HI 1225+01~SW is a real dark galaxy
being witnessed in the process of its merging with the other gas-rich (and
very metal-poor) object.

\section{eXtremely Metal-Deficient (XMD) galaxies: summary of properties}

Due to space limitations, we only give here a very brief summary of XMD
galaxy properties closely related to the further discussion.
The great majority of late-type XMD galaxies known to date (conditionally,
with
metallicities Z of Z$_{\odot}$/34 to Z$_{\odot}$/10, where Z$_{\odot}$
corresponds to 12+log(O/H)=8.66) are classified as blue compact galaxies
(BCGs) which are
low-mass starbursting galaxies. They represent the very edge of the general
BCG metallicity distribution (peaked at Z$\sim$Z$_{\odot}$/5) and comprise
only $\sim$2\% of all known BCGs. Their number known to date is about a
half a hundred.

For ``quiescent'' late-type dwarfs, there exists a well known rather tight
luminosity-metallicity (L--Z) relation (e.g., \cite{Skillman89}), which is
applicable
over $\sim$7 magnitudes in $B$-band and 1.5 dex in O/H. A couple the dimmest
dI galaxies,
UGCA~292 and Leo~A (M$_{\rm B} \sim -11.5$) show Z as low as Z$_{\odot}$/25
(12+log(O/H) $\sim$7.3).
The origin of this L--Z relation is usually explained in terms of a slower
astration in low-mass galaxies and partlially by the elevated metal loss in
the smallest galaxies. A similar L--Z relation for BCGs does exist, albeit
with much larger scattering and with a shift to the higher luminosities at a
fixed O/H. These scattering and shift are especially large in the XMD regime
(see, e.g., a bit out-of-date Figure in \cite{Pustilnik03}).
To emphasize the large difference between XMD dIs and BCGs, we compare their
baryon masses. As follows from Table ~\ref{tab:XMD}, the range of these XMD
BCG baryon masses M$_{\rm bar}$, accepted as M(gas)=M(HI)+M(He), equals
$\sim$(2.6--12)
$\times$10$^8$~M$_{\odot}$. For the most metal-poor dIs Leo~A and UGCA~292,
the M$_{\rm bar}$ are $\sim$0.1 and 0.5$\times$10$^8$~M$_{\odot}$, that is
on average more than an order of magnitude smaller.

The global parameters of XMD BCGs show very large diversity. For their small
metallicity range (a factor of $\sim$3), their L$_{\rm B}$ and
M(HI) vary in the range of 150 and 200, respectively. M(HI)/L$_{\rm B}$
varies between $<$0.2 to 8 (in solar units). For several XMD BCGs the gas
mass-fraction (M$_{\rm gas}$/(M$_{\rm gas}$+M$_{\rm stars}$)) is found to be
as high as 0.95-0.99 (see summary, e.g., in \cite{PM07}).
Morphologies of XMD BCGs vary from regular to typical mergers. All this
implies probable inhomogeniety of XMD BCGs on their evolutionary path-ways.
An additional evidence for this are the colours of their outer parts which
vary from red to very blue (in few galaxies). This implies that the majority
of XMD BCGs are rather old, while a fraction of ``very blue'' gas-dominant XMD
BCGs can be rather ``young'' (namely, their ``first stars'' ages T$* <$ 0.5-2
Gyr $<<$ 13.5 Gyr).

\section{Interactions/mergers in XMD BCGs}
\label{XMD_merger}

The importance of interactions for BCG starbursts in general was
discussed by many authors (e.g., \cite{PKLU01} and references therein).
For XMD BCGs, interaction-induced starbursts are currently known to take place
in at least a half of this group. Curiously enough it appears that all six
of the most metal-poor
BCGs, with 12+log(O/H)=7.12--7.29, show various signs of interactions/mergers.
We summarize their parameters in Table \ref{tab:XMD}.
Due to lack of space we do not show the images with optical/HI morphology and
kinematics. Part of them are published, while the rest will be presented
soon elsewhere. Below we give some notes on these galaxies. The unique merging
XMD galaxy pair SBS 0335--052~E,W with gas mass-fractions of 0.96 and 0.99
provides the best polygon to confront models of very gas-rich mergers with
real objects. The existence of this and another merging system HI 1225+01, in
which the NE component is also an XMD galaxy and the SW component is a
``dark galaxy'' candidate, suggests that there are ``special'' space regions
in which such atypical objects are more abundant and, thus, can be
found in a mutual collision. There are indications that XMD galaxies probably
favour the void-like environment (e.g., \cite{DDO68,HS2134,IAUS235}).
Since the galaxy number density in voids is higher near the borders, one can
expect to detect such objects in transition layers of void regions.

\begin{table}\def~{\hphantom{0}}
  \begin{center}
  \caption{Main parameters of six the most metal-poor BCGs}
  \label{tab:XMD}
  \begin{tabular}{llccccc}\hline
Name            & O/H  & M(HI)&M$_{\rm B}$& V$_{\rm rot}$ & Dist & M$_{\rm bar}$/M$_{\rm tot}$ \\\hline
SBS~0335--052~W & 7.12 & ~9.0 & $-14.7$   & ~37           & ~53  & ~0.2                        \\
DDO~68          & 7.14 & ~7.0 & $-15.5$   & ~51           & ~10  & ~0.2                        \\
I~Zw~18         & 7.17 & ~2.5 & $-15.2$   & ~44           & ~18  & ~0.1                        \\
UGC 772         & 7.24 & ~2.4 & $-14.4$   & ~20           & ~14  &                             \\
SDSS J2134--0035& 7.26 & ~2.0 & $-13.8$   & ~58           & ~20  & ~0.7                        \\
SBS~0335--052~E & 7.29 & ~8.0 & $-16.9$   & ~32           & ~53  & ~0.2                        \\ \hline
  \end{tabular}
 \end{center}
O/H - in units 12+log(O/H) from \cite{Izotov05,Izotov06} and \cite{Izotov07},
M(HI) - in units 10$^8$~M$_{\odot}$ from \cite{Pustilnik_VLA,DDO68,vZee98},
\cite{Schneider91}, and data in preparation, V$_{\rm rot}$ in  km~s$^{-1}$,
Distance in Mpc. V$_{\rm rot}$ for SBS~0335--052~E,W and UGC 772 are lower
and M$_{\rm bar}$/M$_{\rm tot}$
are upper limits, since the inclination correction is uknown.
\end{table}

Summarising, we conclude that previous (e.g., \cite{vZee98},
\cite{Pustilnik_VLA}, \cite{DDO68} for I~Zw~18, SBS 0335--052~E,W and DDO~68)
and new (Ekta et al., in prep. for UGC~772 and
SDSS J2134--0035) observational data indicate that starbursts in all
the lowest metallicity (O/H $<$ 7.30) XMD BCGs, which are ``massive''
objects with M(gas) of (2.6--12)$\times$10$^8$ M$_{\odot}$, are related to
mergers and strong interactions. Their ``high'' baryon and total masses imply
that the metal loss due to galactic winds does not affect their chemical
evolution. Therefore, their extremely low metallicities suggest that their
progenitors are very
stable and have produced very few stars/metals (if any at all) in previous
epochs.
Hence, they should be either very low surface brightness galaxies or a kind
of protogalaxies, which escaped strong external disturbance. If some of the
XMD
BCGs are indeed ``young'', then dark galaxies are the natural candidates to be
their progenitors. They could become ``visible'' due to recent strong
interactions.

\section{Dark Galaxy collisions: an empirical approach and need for models}

It is evident that dark galaxies (DG) are treated as superstable against gas
collapse only if taken as ``isolated'' objects (e.i. when the external
perturbations are smaller than internal ones). Collisions of a fraction of
DGs with galaxy-sized objects should affect their stability and induce the
sinking of gas to their centers and its collapse. One can suggest three
empirical levels of disturbance due to gravitational interaction:
{\it significant, strong} and {\it merger}. We call collision/interaction
``significant'' if it triggers gas collapse and SF and elevates the stellar
mass and related (central) surface brightness above the ``threshold'' level,
say of $\mu =$26.5~$B$-mag~sq.arcsec$^{-2}$, which is characteristic of the
extremely LSB galaxies. How much such an object would resemble known ELSBGs,
depends on its SB radial profile.
For the case of ``strong'' collision/interaction, DG will transform to an
object with a more typical central SB, say with
$\mu =$23--25~$B$-mag~sq.arcsec$^{-2}$.
Their resemblence to known LSBGs again depends on the resulting SB radial
profile. In the case of a merger of a DG and another galaxy-sized aggregate,
the results can be rather different depending on the type and mass
of the said counterpart. Such an event can be accompanied by a significant starburst
and for some cases can look like a XMD BCG.

It is interesting to note that result of ``significant'' interaction can be
transient if a DG keeps its internal stability after the collision. After
0.5--1 Gyr all massive and intermediate mass newly formed stars will die and
the light of this ``ELSB'' galaxy will fade below $\mu =$
27.5~$B$-mag~sq.arcsec$^{-2}$. The object again will be transformed to a dark
galaxy.

This discussion of various galaxies, which could be in principle
related to a population of dark galaxies in the local Universe, shows a
serious need for numerical models of dark galaxy interactions. While the
simulations of very gas-rich galaxy collisions have been difficult until
recently
due to the problem of proper accounting for various feedback processes,
there has been significant progress made over the last two years. \cite{SH05} and
\cite{Robertson06} presented N-body simulations of interacting gas-rich
galaxies (with 99\% of baryons in gas) which reproduce the formation of
a disk galaxy in a major merger. Up to now the models have dealt with rather
massive objects. There is a need to extend them to the region of expected
``dark galaxy'' parametric space. This will allow one to better understand
what emerges from their interactions: more or less ``typical''
LSBGs or something unusual. The models of ``dark galaxy'' mergers will
elucidate whether some of XMD BCGs may be related to this process.

\begin{acknowledgments}

I would like to thank my collaborators A.~Kniazev, A.~Moiseev, J.-M.~Martin,
J.~Chengalur, Ekta, A.~Pramskij, L.~Vanzi and A.~Tepliakova, with whom the
recent results on XMD BCGs have been obtained. I acknowledge partial support
by the IAU travel grant and by RFBR under grant No. 06-02-16617.

\end{acknowledgments}

\end{document}